\documentclass[
 reprint,
 amsmath,amssymb,
 aps,
pra,
]{revtex4-2}
\usepackage{graphicx}
\usepackage{float}% Include figure files
\usepackage{dcolumn}% Align table columns on decimal point
\usepackage{bm}% bold math
\usepackage{bbm}% allows \mathbbm{1} to get 'blackboard' style '1'.
\usepackage{braket}
\usepackage{xcolor}
\usepackage[T1]{fontenc}
\usepackage{enumerate,enumitem}
\usepackage{physics}
\usepackage{xr}

\makeatletter
\DeclareFontFamily{OMX}{MnSymbolE}{}
\DeclareSymbolFont{MnLargeSymbols}{OMX}{MnSymbolE}{m}{n}
\SetSymbolFont{MnLargeSymbols}{bold}{OMX}{MnSymbolE}{b}{n}
\DeclareFontShape{OMX}{MnSymbolE}{m}{n}{
    <-6>  MnSymbolE5
   <6-7>  MnSymbolE6
   <7-8>  MnSymbolE7
   <8-9>  MnSymbolE8
   <9-10> MnSymbolE9
  <10-12> MnSymbolE10
  <12->   MnSymbolE12
}{}
\DeclareFontShape{OMX}{MnSymbolE}{b}{n}{
    <-6>  MnSymbolE-Bold5
   <6-7>  MnSymbolE-Bold6
   <7-8>  MnSymbolE-Bold7
   <8-9>  MnSymbolE-Bold8
   <9-10> MnSymbolE-Bold9
  <10-12> MnSymbolE-Bold10
  <12->   MnSymbolE-Bold12
}{}

\let\llangle\@undefined
\let\rrangle\@undefined
\DeclareMathDelimiter{\llangle}{\mathopen}%
                     {MnLargeSymbols}{'164}{MnLargeSymbols}{'164}
\DeclareMathDelimiter{\rrangle}{\mathclose}%
                     {MnLargeSymbols}{'171}{MnLargeSymbols}{'171}
\makeatother

\begin{document}

\title{Homodyne detection is optimal for quantum interferometry \\with path-entangled coherent states}

\author{Z.~M.~McIntyre }
\email{zoe.mcintyre@mail.mcgill.ca}
\author{W.~A.~Coish}%
 \email{william.coish@mcgill.ca}
\affiliation{%
 Department of Physics, McGill University, 3600 rue University, Montreal, QC, H3A 2T8, Canada
}%

\date{\today}

\begin{abstract}
   We present measurement schemes that do not rely on photon-number resolving detectors, but that are nevertheless optimal for estimating a differential phase shift in interferometry with either an entangled coherent state or a qubit--which-path state (where the path taken by a coherent-state wavepacket is entangled with the state of a qubit). The homodyning schemes analyzed here achieve optimality (saturate the quantum Cram\'er-Rao bound) by maximizing the sensitivity of  measurement outcomes to phase-dependent interference fringes in a reduced Wigner distribution. In the presence of photon loss, the schemes become suboptimal, but we find that their performance is independent of the phase to be measured. They can therefore be implemented without any prior information about the phase and without adapting the strategy during measurement, unlike strategies based on photon-number parity measurements or direct photon counting.
\end{abstract}

\maketitle

Interferometry for phase estimation is one of the fundamental tasks of quantum metrology~\cite{ye2024essay}, with applications in fields ranging from biophysics~\cite{taylor2016quantum,moreau2019imaging,mukamel2020roadmap} to gravitational wave detection~\cite{aasi2013enhanced,tse2019quantum,yu2020quantum}. The ultimate goal of quantum-enhanced interferometry is to determine an unknown phase $\phi$ with a precision better than the standard quantum limit (shot-noise limit) for uncorrelated photons, given by $\Delta \phi\geq\delta\phi_\textsc{sql}= N^{-1/2}$, where $\Delta \phi$ is the standard deviation of $\phi$ and $N$ is the number of photons that pass through the interferometer in a single measurement.  Correlations arising from non-classical states can, however, lead to better phase sensitivity, with improved scaling at the Heisenberg limit, $\Delta\phi\propto N^{-1}$. 

\begin{figure}
    \centering
    \includegraphics[width=0.47\textwidth]{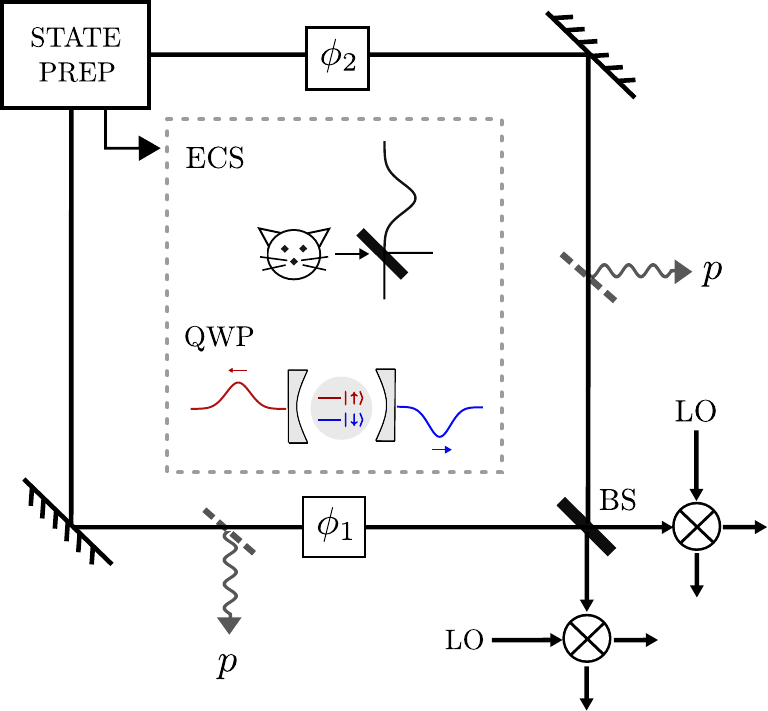}
    \caption{A Mach-Zehnder interferometer can be used to estimate the differential phase shift $\phi=\phi_1-\phi_2$. Photon loss from the interferometer occurs with a probability-per-photon $p$. Homodyne detection is implemented by mixing the output of the interferometer with a local oscillator (LO) field. State preparation: An ECS can be produced by mixing an even cat state $\propto \ket*{\alpha/\sqrt{2}}+\ket*{-\alpha/\sqrt{2}}$ with a coherent state $\ket*{\alpha/\sqrt{2}}$ on a 50:50 beamsplitter~\cite{gerry2009maximally,joo2011quantum}. In a cavity-QED setup, a QWP state can be generated by feeding a coherent-state wavepacket into the input port of a cavity containing a qubit prepared in $\ket{+}\propto\ket{\uparrow}+\ket{\downarrow}$, while also modulating the strength of an asymmetric longitudinal ($\propto \ketbra{\uparrow}$) cavity-qubit coupling~\cite{mcintyre2024photonic}.}
    \label{fig:interferometer}
\end{figure}

The first study of quantum-enhanced interferometry considered phase estimation with a Mach-Zehnder interferometer (Fig.~\ref{fig:interferometer}) fed by a coherent state mixed on a beamsplitter with a squeezed vacuum state~\cite{caves1981quantum}. In this configuration, Heisenberg-limited precision can be achieved by counting the precise number of photons arriving at each of two output ports of the interferometer~\cite{pezze2008mach,lang2013optimal}. Photon counting is in fact optimal for this state, in the sense that it enables the best precision allowed by quantum mechanics, $\delta\phi_{\mathrm{min}}$,  given by the quantum Cram\'er-Rao bound (CRB)~\cite{helstrom1967minimum,braunstein1994statistical,braunstein1996generalized},
\begin{equation}\label{qcrb}
    \Delta\phi\geq\delta\phi_{\mathrm{min}}= \frac{1}{\sqrt{MI_{\mathrm{Q}}(\rho_\phi)}},
\end{equation}
where here, $M$ is the number of independent measurements and $I_{\mathrm{Q}}(\rho_\phi)$ is the quantum Fisher information of $\rho_\phi=e^{-i\phi A}\rho(0)e^{i\phi A}$ with respect to $A$, the generator of $\phi$. Formally, the quantum Fisher information is given by $I_{\mathrm{Q}}(\rho_\phi)=\mathrm{Tr}\{\rho_\phi\mathcal{L}^2\}$, with the symmetric logarithmic derivative operator $\mathcal{L}$ defined implicitly through the relation $\partial_\phi\rho_\phi=(\mathcal{L}\rho_\phi+\rho_\phi\mathcal{L})/2$~\cite{toth2014quantum}.

Photon counting provides an optimal strategy not just for a coherent state mixed with squeezed vacuum~\cite{pezze2008mach, lang2013optimal}, but for any path-symmetric pure state~\cite{hofmann2009all}. The class of path-symmetric states includes many of the states most commonly considered for quantum metrology, such as N00N states~\cite{bollinger1996optimal,gerry2000heisenberg,gerry2002nonlinear}, twin Fock states~\cite{holland1993interferometric}, two-mode squeezed vacuum states~\cite{anisimov2010quantum}, and entangled coherent states (ECS's)~\cite{joo2011quantum}. This optimal measurement strategy may, however, be associated with additional technological complexity: Photon-number resolving detectors (typically, superconducting transition-edge sensors) must be kept at cryogenic temperatures, and state-of-the-art number-resolving detectors have only now demonstrated the ability to resolve up to $\sim 100$ photons~\cite{eaton2023resolution}, while phase-sensitive (quadrature) measurements like homodyne and heterodyne detection require less-specialized equipment.

In this Letter, we present homodyne-detection-based schemes that are optimal (in the absence of photon loss, $p=0$ in Fig.~\ref{fig:interferometer}) for quantum interferometry with either of two path-entangled coherent states---an ECS~\cite{sanders1992entangled} or a qubit--which-path (QWP) state~\cite{mcintyre2024photonic}:
\begin{align}
    &\ket{\mathrm{ECS}}=\mathcal{N}_\alpha(\ket{\alpha,0}+\ket{0,\alpha}),\label{ECS}\\
    &\ket{\mathrm{QWP}}=\frac{1}{\sqrt{2}}\left(\ket{\uparrow}\ket{\alpha,0}+\ket{\downarrow}\ket{0,\alpha}\right),\label{QWP}
\end{align}
where $\mathcal{N}_\alpha=[2(1+e^{-\lvert\alpha\rvert^2})]^{-1/2}$. Here, $\ket{\alpha_1,\alpha_2}=\prod_{i=1,2}\mathrm{D}_i(\alpha_i)\ket{0}$, with vacuum state $\ket{0}$ and displacement operator $\mathrm{D}_i(\alpha)=e^{\alpha a_i^\dagger-\mathrm{h.c.}}$. This is a two-mode coherent state with amplitude $\alpha_i$ in the traveling-wave mode that is annihilated by $a_i$, located in arm $i=1,2$ of the interferometer. In Eq.~\eqref{QWP}, the states $\ket{\uparrow},\ket{\downarrow}$ are energy eigenstates of a two-level system (qubit). 

As light passes through the interferometer, the initially prepared state acquires a dependence on the differential phase $\phi=\phi_1-\phi_2$ through unitary evolution generated by $U_{\phi}=\prod_{i=1,2}e^{-i\phi_in_i}$, where $n_i=a_i^\dagger a_i$ is the number operator for mode $i$. For a pure state, the quantum Fisher information $I_{\mathrm{Q}}$ of $\ket{S_\phi}=U_{\phi}\ket{S}$ is given in terms of the variance of $J_3=(n_1-n_2)/2$ with respect to $\ket{S_\phi}$ as $I_{\mathrm{Q}}(\ket{S_\phi})=4\mathrm{Var}_{\ket{S_\phi}}(J_3)$~\cite{toth2014quantum}. Evaluating the variance gives
\begin{align}
    &I_{\mathrm{Q}}(\ket{\mathrm{ECS}_\phi})=\bar{n}^2+[1+w(\bar{n}e^{-\bar{n}})]\bar{n},\label{quantum-fisher-ECS}\\
    &I_{\mathrm{Q}}(\ket{\mathrm{QWP}_\phi})=\bar{n}^2+\bar{n},\label{quantum-fisher-QWP}
\end{align}
where $\bar{n}=\langle n_1+n_2\rangle$ is the total average number of photons, and where $w(z)$ is the Lambert $W$ function. For a QWP state, $\bar{n}=\lvert\alpha\rvert^2$. For an ECS, however, $\bar{n}=\lvert\alpha\rvert^2/(1+e^{-\lvert\alpha\rvert^2})$. Inverting this relation is what produces a dependence on $w(\bar{n}e^{-\bar{n}})$. The term $\propto w$ in  $I_{\mathrm{Q}}(\ket{\mathrm{ECS}_\phi})$ provides a small advantage over the QWP state at small $\bar{n}$. For large $\bar{n}$, however, the advantage is exponentially suppressed since $w(\bar{n}e^{-\bar{n}})\simeq \bar{n}e^{-\bar{n}}$ for $\bar{n}\gg 1$. At large $\bar{n}$, both ECS's and QWP states provide Heisenberg-limited scaling $\propto\bar{n}^2$. Both states (ECS and QWP) also have a small precision advantage over N00N states consisting of superpositions $\ket{N00N}\propto \ket{N,0}+\ket{0,N}$ of $N$-photon Fock states, for which $I_{\mathrm{Q}}(\ket{N00N_\phi})=N^2$~\cite{bollinger1996optimal,gerry2000heisenberg,gerry2002nonlinear}. An analogous expression for the quantum Fisher information of an ECS [Eq.~\eqref{quantum-fisher-ECS}] was derived in Ref.~\cite{zhang2013quantum} for estimation of the total phase shift $\phi_1$ in mode 1 (generated by $a_1^\dagger a_1$), rather than estimation of the differential phase shift $\phi=\phi_1-\phi_2$ (generated by $J_3$). 

Not every measurement scheme can be used to saturate the quantum CRB. For a scheme where $\phi$ is estimated by measuring some quantity $\mathcal{O}$ having outcomes $x$, described by the positive-operator valued measure (POVM) $\{\hat{\Pi}_x\}$, the standard deviation $\Delta\phi$ of any unbiased estimator $\hat{\phi}(x)$ has a lower bound given by the classical CRB~\cite{cramer1999mathematical},
\begin{equation}\label{classical-CRB}
   \Delta\phi\geq \delta\phi= \frac{1}{\sqrt{M I_{\mathrm{C}}(\phi)}}.
\end{equation}
Here, the classical Fisher information $I_{\mathrm{C}}(\phi)$ is given by 
\begin{equation}
    I_{\mathrm{C}}(\phi)\equiv I_{\mathrm{C}}[p(x\vert\phi)]=\int dx\left(\partial_\phi \mathrm{ln}\:p(x\vert\phi)\right)^2p(x\vert \phi),\label{classical-fisher}
\end{equation}
where $p(x\lvert\phi)=\mathrm{Tr}\{\rho_\phi \hat{\Pi}_x\}$. Under some regularity conditions [requiring, for instance, that $p(x\lvert\phi)$ have a unique global maximum], the maximum-likelihood estimator $\hat{\phi}_{\textsc{mle}}(\bm{x})=\mathrm{argmax}_\phi p(\bm{x}\vert \phi)$ saturates the CRB in the asymptotic limit $M\rightarrow \infty$, where here, $\bm{x}=\{x_i\}_{i=1}^M$ is a set of observations sampled from $p(x\vert\phi)$~\cite{cramer1999mathematical}. Since the classical CRB can be saturated in principle, a measurement scheme is optimal when its classical Fisher information $I_{\mathrm{C}}(\phi)$ is equal to $I_{\mathrm{Q}}(\rho_\phi)$, in which case $\delta\phi=\delta\phi_{\mathrm{min}}$.

\begin{figure}
    \centering
    \includegraphics[width=\columnwidth]{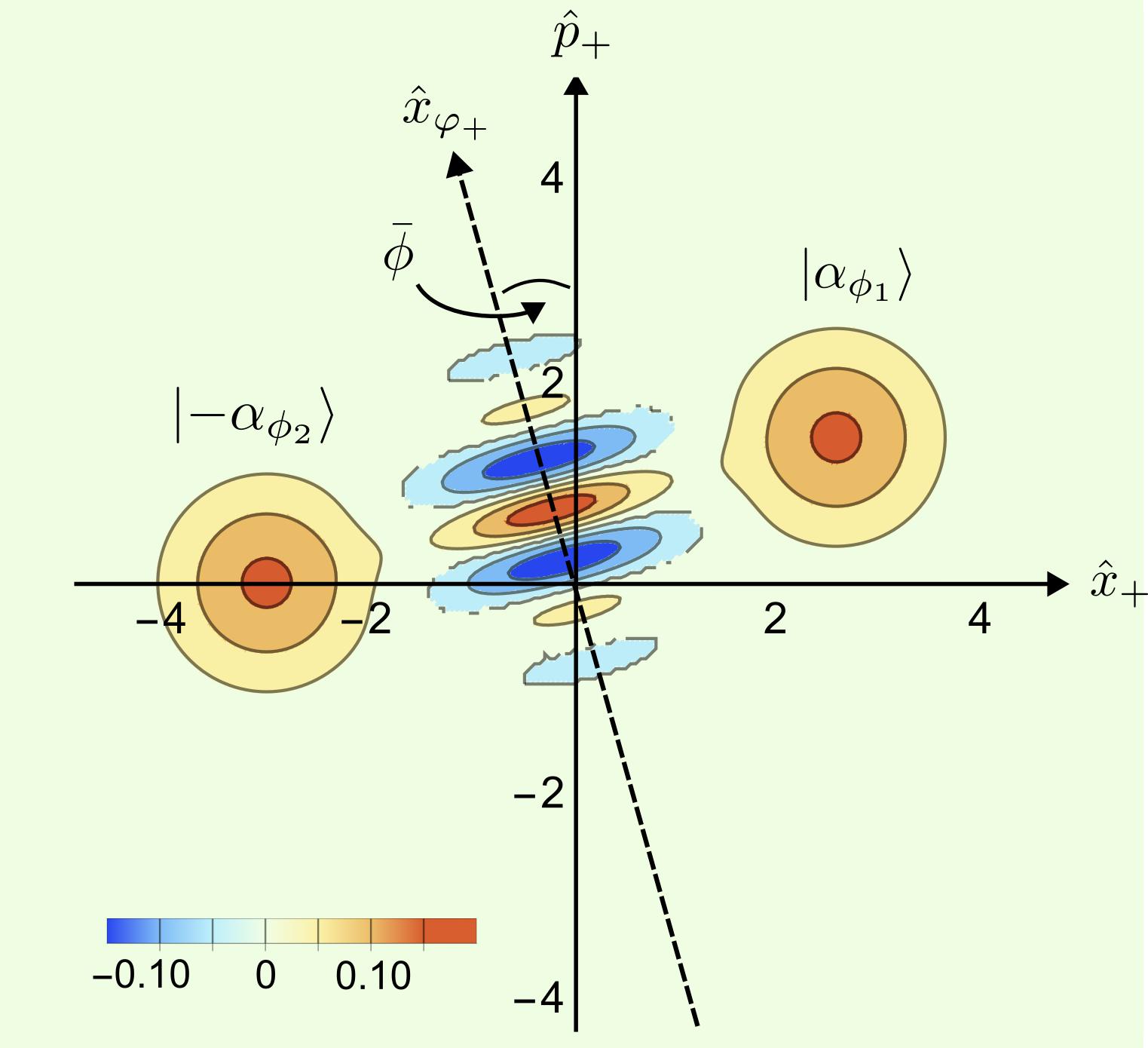}
    \caption{Reduced Wigner distribution $W(x_+,p_+)=\int dx_-dp_- W(x_+,p_+,x_-,p_-)$ of mode $a_+$, where here, $W(x_+,p_+,x_-,p_-)$ is the Wigner distribution of the state resulting from an initial ECS [Eq.~\eqref{ECS-post-beamsplitter}] for $\alpha=3$, $\phi_1=0.5$, and $\phi_2=0$. Homodyne detection of mode $a_+$ with a local-oscillator phase of $\varphi_+=\pi/2+\bar{\phi}$ implements a projection onto the rotated quadrature indicated by the dashed black line. Here, $\alpha_{\phi_j}=e^{i\phi_j}\alpha/\sqrt{2}$ [cf.~Eqs.~\eqref{ECS-post-beamsplitter} and \eqref{QWP-post-beamsplitter}].}
    \label{fig:homodyne}
\end{figure}

We now explain how homodyne detection can be used to achieve an optimal measurement for ECS's and QWP states in the absence of photon loss. After light passes through the interferometer, the initially prepared state $\ket{S}$ is mapped to $\ket{S_\phi}$. The light is then passed through a 50:50 beamsplitter BS (Fig.~\ref{fig:interferometer}) that maps the interferometer modes $a_i$ ($i=1,2$) to output modes $a_\pm=(a_1\pm a_2)/\sqrt{2}$ via a unitary operation $U_{\mathrm{BS}}$. The resulting state $\ket*{\widetilde{S}_\phi}=U_{\mathrm{BS}}\ket{S_\phi}$ is then given by
\begin{align}
    &\ket*{\widetilde{\mathrm{ECS}}_\phi}=\mathcal{N}_\alpha(\ket*{\alpha_{\phi_1},\alpha_{\phi_1}}+\ket*{-\alpha_{\phi_2},\alpha_{\phi_2}}),\label{ECS-post-beamsplitter}\\
    &\ket*{\widetilde{\mathrm{QWP}}_\phi}=\frac{1}{\sqrt{2}}(\ket{\uparrow}\ket*{\alpha_{\phi_1},\alpha_{\phi_1}}+\ket{\downarrow}\ket*{-\alpha_{\phi_2},\alpha_{\phi_2}}),\label{QWP-post-beamsplitter}
\end{align}
where $\alpha_{\phi_j}=e^{i\phi_j}\alpha/\sqrt{2}$.
The measurement schemes consist of (I) measuring modes $a_\pm$ with homodyne detection using local-oscillator phases $\varphi_\pm$, respectively, where 
\begin{align}\label{local-oscillator-phase}
\begin{aligned}
    &\varphi_+=\frac{\pi}{2}+\bar{\phi},\\
    &\varphi_-=\bar{\phi},\\
    &\bar{\phi}=\frac{1}{2}(\phi_1+\phi_2).
\end{aligned}
\end{align}
Prior information about the average phase $\bar{\phi}$ is therefore required. For the ECS, that completes the measurement. In the case of the QWP state, the homodyne measurements are followed by (II) a measurement of the qubit in the Pauli-$X$ basis, with outcomes $X=\pm$ for states $\ket{\pm}=\left(\ket{\uparrow}\pm\ket{\downarrow}\right)/\sqrt{2}$. 

To evaluate the classical Fisher information [Eq.~\eqref{classical-fisher}] associated with the measurement schemes presented here, we derive conditional probability distributions $p_{S}(x\vert\phi)$ governing the measurement outcomes, where for both states $S=\mathrm{ECS}, \mathrm{QWP}$, the variable $x$ includes the two outcomes for homodyne detection of modes $a_\pm$ [Step (I)], and where for the QWP state, $x$ also includes the outcome of the $X$-basis qubit measurement [Step (II)]. We find that in the absence of photon loss ($p=0$), the measurement schemes described by (I)-(II) are optimal,
\begin{equation}\label{optimal}
    I_{\mathrm{C}}[p_{S}(x\vert\phi)]=I_{\mathrm{Q}}(\ket{S_\phi}),\quad S=\mathrm{ECS,QWP}.
\end{equation}
The optimality is a consequence of choosing local-oscillator phases $\varphi_\pm$ [Eq.~\eqref{local-oscillator-phase}] that make the measurement outcomes maximally sensitive to the $\phi$-dependent fringes in the Wigner distribution of $\ket*{\Tilde{S}_\phi}$ [Eqs.~\eqref{ECS-post-beamsplitter} and \eqref{QWP-post-beamsplitter}] (see Fig.~\ref{fig:homodyne} for the case of $S=\mathrm{ECS}$). The measurement schemes presented above only make use of homodyne detection and (in the case of the QWP state) single-qubit control/readout. Notably, we have found that an optimal measurement for the QWP state can be devised without the use of entangling operations, such as the controlled-phase gate considered in Ref.~\cite{feng2023robust} as a way of mapping phase information from a bosonic system into the state of a qubit. Such entangling operations may be difficult to implement in an interferometer. Additionally, while the authors of Ref.~\cite{knott2014attaining} have argued that achieving Heisenberg-limited metrology with an ECS cannot be accomplished with homodyning, we show here that this is untrue provided we have prior information about $\bar{\phi}$.

\begin{figure}
    \centering
    \includegraphics[width=0.49\textwidth]{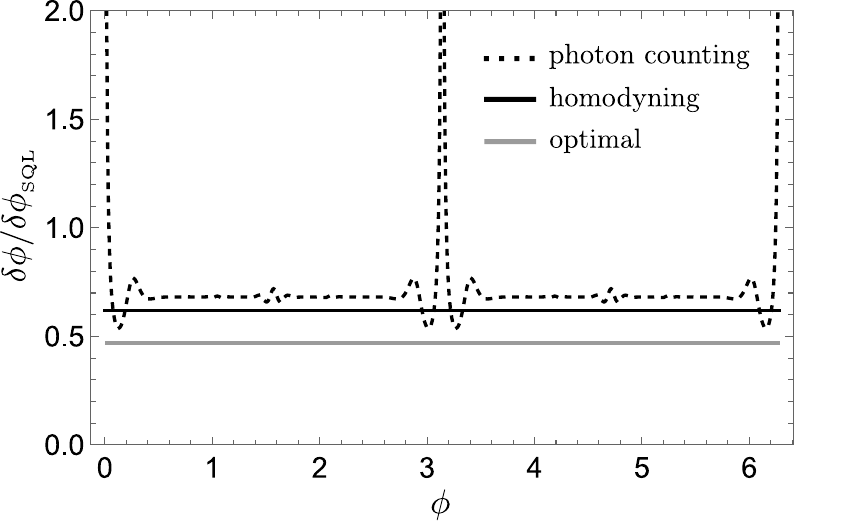}
    \caption{Precision $\delta\phi$ [Eq.~\eqref{classical-CRB}] as a function of $\phi$, relative to the standard quantum limit $\delta\phi_{\textsc{sql}}\equiv [(1-p)M\bar{n}]^{-1/2}$, for an ECS with $\bar{n}=10$ and $p=0.05$ ($5\%$ photon loss). The values that can be achieved with homodyne detection (solid black line) and  photon counting (dashed black line) were calculated using the probability distributions given in Eqs.~\eqref{homodyne-probabilities-photon-loss} and \eqref{photon-counting-probabilities}, respectively. The gray line corresponds to the optimal precision [$\delta\phi_{\mathrm{min}}$ with the quantum Fisher information given in Eq.~\eqref{quantum-fisher-ECS-loss}]. }
    \label{fig:phase-dep}
\end{figure}

A consequence of Eq.~\eqref{optimal} is that the precision $\delta\phi\propto [I_{\mathrm{C}}(\phi)]^{-1/2}$ that can be achieved using these measurements is independent of the true value of $\phi$ (since $I_{\mathrm{Q}}$ is $\phi$-independent), allowing for an optimal non-adaptive measurement without prior knowledge of $\phi$. In the case of an ECS, this can be contrasted to the scheme based on photon-number parity measurements~\cite{joo2011quantum}, where information about $\phi$ is extracted by determining whether the number of photons in one of the output modes of the interferometer is even or odd. Although parity measurements are sub-optimal, they can nevertheless be used to achieve Heisenberg-limited scaling~\cite{joo2011quantum}. For $\ket*{\widetilde{\mathrm{ECS}_\phi}}$, the probability $p(\mathrm{even}\vert\phi)$ of measuring an even number of photons in one of the output modes exhibits $\phi$-dependent oscillations that can be used to extract information about $\phi$~\cite{joo2011quantum}. However, since the visibility of these oscillations is suppressed by a factor $e^{-\lvert\alpha\rvert^2\mathrm{sin}^2(\phi/2)}$, the scheme is effective in the limit $\bar{n}\simeq|\alpha|^2\gg 1$ only if $\lvert\alpha\rvert^2\phi^2\ll 1$, requiring prior knowledge of $\phi$ with a precision $\sim 1/\lvert\alpha\rvert$. The need for prior characterization of $\phi$ could be eliminated by retaining the full counting statistics, as photon counting is also optimal for an ECS~\cite{hofmann2009all,lee2016quantum}. (We find that photon counting is optimal for a QWP state as well, when supplemented by a final $X$-basis measurement of the qubit \cite{supplement}.) However, as soon as photon loss is introduced, the classical Fisher information associated with photon counting acquires a dependence on $\phi$, and some amount of \emph{a priori} knowledge is required in order to avoid values of $\phi$ where the Fisher information vanishes (in which case $\delta\phi\rightarrow\infty$). As we now show, the homodyne-based measurement schemes presented here do not suffer from this drawback (Fig.~\ref{fig:phase-dep}).

From this point onward, we focus on the ECS and therefore dispense with the use of explicit subscripts indicating the state being considered. The results for the QWP state are qualitatively similar and are given in the Supplementary Material~\cite{supplement}.  

To investigate performance accounting for photon loss, we model losses in the interferometer by inserting a fictitious beamsplitter into each interferometer arm~\cite{zhang2013quantum}. These beamsplitters are modeled by the operator $R_{c,c_\ell}(p)=e^{\mathrm{arcsin}\sqrt{p}(c_\ell^\dagger c-\mathrm{h.c.})}$, describing scattering of photons from mode $c$ into loss mode $c_\ell$ with probability $p$. Under the action of the lossy interferometer, the initial state $\ket{\mathrm{ECS}}$ [Eq.~\eqref{ECS}]  evolves to \begin{equation}
    \rho_\phi=\mathrm{Tr}_{\ell}\{RU_{\phi}\rho_{0}U_\phi^\dagger R^\dagger\},\quad R=\prod_{i=1,2}R_{a_i,a_{\ell_i}}(p),\label{rho-phi-loss}
\end{equation}
where $\rho_0=\ketbra{\mathrm{ECS}}\otimes\ketbra{0}_{\ell}$ is the initial state of the interferometer and loss modes (annihilated by $a_{\ell_i}$, $i=1,2$), and where $\mathrm{Tr}_{\ell}$ describes a trace over the state of both loss modes. Note that the same state $\rho_\phi$ is obtained regardless of the order in which $U_{\phi}$ and $R$ are applied. For a mixed state $\rho_\phi$, the quantum Fisher information of $\rho_\phi$ with respect to $J_3$ can be calculated by evaluating matrix elements of $J_3$ in the eigenbasis of $\rho_\phi$~\cite{toth2014quantum}. This procedure gives
\begin{align}
\begin{aligned}
    I_{\mathrm{Q}}(\rho_\phi)&=(1-p)^2\bar{n}^2e^{-2p[\bar{n}+w(\bar{n}e^{-\bar{n}})]}\\
    &+(1-p)\bar{n}[1+(1-p)w(\bar{n}e^{-\bar{n}})],\label{quantum-fisher-ECS-loss}\\
\end{aligned}
\end{align}
where $w(z)$ is again the Lambert $W$ function. Photon loss therefore controls a transition from Heisenberg-limited ($\propto\bar{n}^2$) scaling to scaling at the standard quantum limit ($\propto\bar{n}$). An analogous result for estimation of the total phase shift $\phi_1$ in arm 1 (rather than $\phi=\phi_1-\phi_2$), accounting for photon loss, was presented in Ref.~\cite{zhang2013quantum}. A detailed derivation of the quantum Fisher information of the QWP state, accounting for photon loss and qubit dephasing, is given in the Supplementary Material~\cite{supplement}.

Homodyne detection is performed by mixing the signal field with a local oscillator prepared in a coherent state $\ket{\beta}$, where here, we assume that $\beta\in\mathbb{R}^+$. In the strong-oscillator limit $\lvert \beta\rvert\gg  \lvert\alpha\rvert$, homodyne detection of mode $a$ with a local oscillator in state $\ket*{\beta e^{i(\varphi-\pi)}}$ implements a projection onto the eigenbasis $\ket{x_\varphi}=e^{-i\varphi a^\dagger a}\ket{x}$ of the rotated quadrature operator $\hat{x}_\varphi=\hat{x}\cos{\varphi}+\hat{p}\sin{\varphi}$~\cite{tyc2004operational}, where here, $\hat{x}=(a^\dagger+a)/\sqrt{2}$ and $\hat{p}=i(a^\dagger-a)/\sqrt{2}$ are canonically conjugate, and where $\ket{x}$ is an eigenstate of $\hat{x}$ with eigenvalue $x$. For measurement of mode $a_+$ with local-oscillator phase $\varphi_+=\pi/2+\bar{\phi}$ [Eq.~\eqref{local-oscillator-phase}], this corresponds to projecting the coherent state in mode $a_+$ onto a quadrature rotated by an amount $\bar{\phi}$ relative to the out-of-phase quadrature: $\hat{x}_{\varphi_+}=-\hat{x}_+\sin{\bar{\phi}}+\hat{p}_+\cos{\bar{\phi}}$ (Fig.~\ref{fig:homodyne}). For the measurement scheme presented here, the POVM element describing the measurement of the ECS [Step (I)] is therefore given by $\hat{\Pi}_x=\bigotimes_{\sigma=\pm}e^{-i\varphi_\sigma a_\sigma^\dagger a_\sigma}\ketbra{x_\sigma}e^{i\varphi_\sigma a_\sigma^\dagger a_\sigma}$. Without loss of generality, we assume that $\alpha\in\mathbb{R}$, in which case the probability distribution $p(x\lvert\phi)=\mathrm{Tr}\{\rho_\phi\hat{\Pi}_x\}$ governing the homodyne-measurement outcomes is given by
\begin{align}\label{homodyne-probabilities-photon-loss}
\begin{aligned}
    p(x\lvert\phi)&=2\mathcal{N}_\alpha^2\left[1+e^{-p\alpha^2}\cos{\Theta_{x}(\phi)}\right]\prod_{s=\pm}g_{s}(x_s,\phi),
\end{aligned}
\end{align}
where $g_s(x_s,\phi)=\pi^{-1/2}\exp{{-}[x_s-\mu_s(\phi)]^2}$, $\mu_+(\phi)=\sqrt{1-p}\alpha\sin{\frac{\phi}{2}}$, $\mu_-(\phi)=\sqrt{1-p}\alpha\cos{\frac{\phi}{2}}$, and $\Theta_x(\phi)=2x_+\mu_-(\phi)-2x_-\mu_+(\phi)$. Setting $p=0$, this result [Eq.~\eqref{homodyne-probabilities-photon-loss}] recovers Eq.~\eqref{optimal} for $S=\mathrm{ECS}$.

The term $\sim\cos{\Theta_x(\phi)}$ in Eq.~\eqref{homodyne-probabilities-photon-loss} is a consequence of phase-space interference in the Wigner distribution of $\rho_\phi$. To build intuition for this, consider the single-mode cat state $\ket{\mathrm{C}_+}\propto(\ket{\alpha}+\ket{-\alpha})$. For $\alpha\in\mathbb{R}$, the states $\ket{\pm\alpha}$ are displaced along the $\hat{x}$ quadrature. A homodyne measurement with a local-oscillator phase $\pi/2$ (corresponding to a projection onto the $\hat{p}$ axis) then returns a displacement $x_{\pi/2}$ with probability $p(x_{\pi/2})\propto (1+\cos{\sqrt{8}\alpha x_{\pi/2}})$~\cite{dragan2001homodyne}, where here, the oscillating term is a reflection of interference fringes parallel to the $\hat{x}$-axis in the Wigner distribution of $\ket{\mathrm{C}_+}$. For the measurement of modes $a_\pm$ proposed here, the local oscillator phases $\varphi_\pm$ [Eq.~\eqref{local-oscillator-phase}] are both chosen so that the phase-space axis associated with the measured quadrature $\hat{x}_{\varphi_\pm}$ bisects the angle subtended by the coherent-state displacement of modes $a_\pm$ in the two branches of $\ket*{\widetilde{\mathrm{ECS}}_\phi}$ (Fig.~\ref{fig:homodyne}). Measurements of displacements along these axes are therefore maximally sensitive to the interference fringes between the two branches, resulting in an optimal detection scheme in the ideal scenario of zero photon loss ($p=0$) [cf.~Eq.~\eqref{optimal}]. The dependence of these interference fringes on $\phi$ is what produces Heisenberg-limited scaling $\propto\bar{n}^2$ in the classical Fisher information for this measurement scheme.

\begin{figure}
    \centering
    \includegraphics[width=0.49\textwidth]{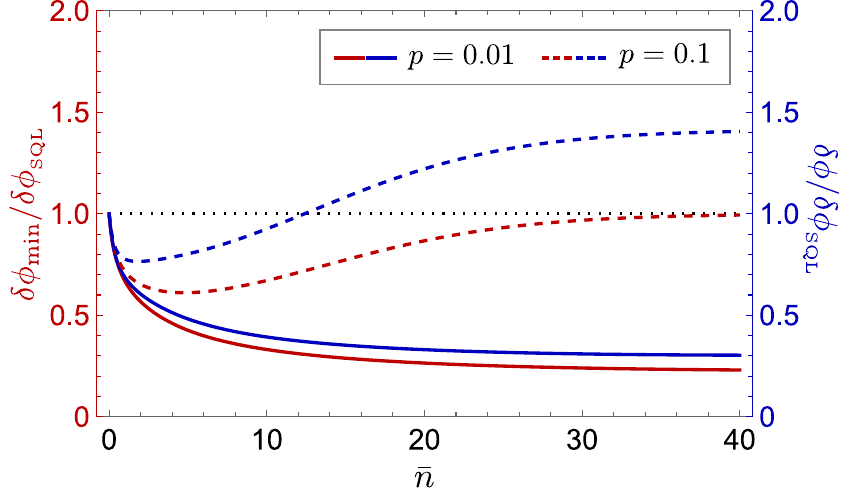}
    \caption{Left axis: Precision given by the quantum CRB, $\delta\phi_{\mathrm{min}}$, relative to the standard quantum limit $\delta\phi_{\textsc{sql}}\equiv[(1-p)M\bar{n}]^{-1/2}$, for an ECS with photon loss $p=0.01$ (solid red line) and $p=0.1$ (dashed red line). Right axis: Precision that can be attained with homodyning, $\delta\phi$, for $p=0.01$ (solid blue line) and $p=0.1$ (dashed blue line).  }
    \label{fig:precision-ECS}
\end{figure}

In Fig.~\ref{fig:precision-ECS}, we compare the precision $\delta\phi$ that can be achieved using this homodyning scheme to $\delta\phi_{\mathrm{min}}$ [Eq.~\eqref{qcrb}] for two values of $p$. For $\bar{n}\gg p^{-1}$, the performance of the homodyning scheme saturates at $\delta\phi=\sqrt{2}\delta\phi_{\textsc{sql}}$ (Fig.~\ref{fig:precision-ECS}). This is because for $\bar{n}\simeq \alpha^2\gg p^{-1}$, the interference term in Eq.~\eqref{homodyne-probabilities-photon-loss} is exponentially suppressed, and $p(x\vert\phi)$ is given approximately by the product of two Gaussians: $p(x\vert\phi)\approx \prod_{s=\pm}g_s(x_s,\phi)$. In this case, $I_{\mathrm{C}}[p(x\vert\phi)]\approx I_{\mathrm
C}[g_+(x_+,\phi)]+I_{\mathrm
C}[g_-(x_-,\phi)]$. Noting that $\mu_\pm(\phi)$ both oscillate with a period $4\pi$ (rather than $2\pi$), the factor of $\sqrt{2}$ relating $\delta\phi$ to $\delta\phi_{\textsc{sql}}$ in the limit $p\bar{n}\gg1 $ can therefore be understood as a consequence of ``sub-resolution'' in the Gaussian distributions, to be contrasted with super-resolution~\cite{resch2007time}, where the distributions would instead depend on an amplified phase $m\phi$ with $m>1$.

As discussed above, photon counting can also be used to saturate the quantum CRB for an ECS in the absence of photon loss~\cite{hofmann2009all,lee2016quantum}. Accounting for photon loss, the probability $p(m,n\lvert\phi)$ of detecting $m$ and $n$ photons in modes $a_+$ and $a_-$, respectively, is given by
\begin{align}\label{photon-counting-probabilities}
\begin{aligned}
    p(m,n\lvert\phi)&=2\mathcal{N}_\alpha^2\left[1+e^{-p\alpha^2}\cos{\Theta_{m,n}}(\phi)\right]\prod_{j=m,n}P(j;\lambda_\alpha),
\end{aligned}
\end{align}
where $\Theta_{m,n}(\phi)=(m+n)\phi+m\pi$, $P(j;\lambda)=e^{-\lambda}\lambda^j/j!$, and $\lambda_\alpha=(1-p)\alpha^2/2$. We have verified numerically that the Fisher information $I_{\mathrm{C}}[p(x\vert\phi)]$ is independent of $\phi$, while $I_{\mathrm{C}}[p(m,n\vert\phi)]=0$ for $\phi=0,\pi$, leading to singularities in Fig.~\ref{fig:phase-dep}. In the asymptotic limit $M\rightarrow \infty$, the distribution of outcomes $(\hat{\phi}_{\textsc{mle}}-\phi)$ associated with the maximum-likelihood estimate $\hat{\phi}_{\textsc{mle}}$ of $\phi$ converges to a zero-mean normal distribution with variance $\delta\phi^2=1/MI_{\mathrm{C}}(\phi)$~\cite{cramer1999mathematical}. For maximum-likelihood estimation in the vicinity of $\phi=0,\pi$, however, the maximum-likelihood estimator will not converge to the true value of $\phi$ when an estimation strategy based on photon counting is used. This caveat is not present when the homodyning scheme is used instead, due to the phase independence of $I_{\mathrm{C}}[p(x\vert\phi)]$ (Fig.~\ref{fig:phase-dep}).

Here, we have presented measurement schemes based on homodyne detection that are optimal, in the absence of photon loss, for interferometry using either an ECS or a QWP state. The schemes achieve optimality by using prior knowledge of the average phase $\bar{\phi}$ to choose local-oscillator phases that maximize the sensitivity of measurement outcomes to the $\phi$-dependent interference fringes in the states' Wigner distributions. We have also shown that the achievable precision, as given by the CRB, is independent of the true value of $\phi$, even in the presence of photon loss. 

A natural extension of the strategies used here would be to investigate whether an optimal homodyning scheme can be found for the Caves state (produced by mixing a coherent state with squeezed vacuum \cite{caves1981quantum}). For a coherent state combined on a beamsplitter with any other quantum state of light, it was found that the squeezed vacuum produces the largest quantum Fisher information at a fixed average photon number \cite{lang2013optimal}. A Caves state therefore has greater potential sensitivity than either the ECS or the QWP state investigated here. For a Caves state, it is known that photon-number parity measurements saturate the quantum CRB in the absence of photon loss, but only in the vicinity of $\phi=0$ \cite{seshadreesan2011parity}. In the presence of photon loss, it was found in Ref.~\cite{gard2017nearly} that a non-optimal homodyning scheme for the Caves state exhibited better sensitivity than parity measurements. Optimizing the homodyning scheme for this state using the ideas presented here could therefore lead to a better and more practical inference method.

\nocite{norris2016qubit,paz2017multiqubit,kwiatkowski2020influence,mcintyre2022non}

\textit{Acknowledgements}---We thank J.~Sankey for useful discussions. We also acknowledge funding from the Natural Sciences and Engineering Research Council of Canada (NSERC) and from the Fonds de Recherche du Qu\'ebec--Nature et technologies (FRQNT).

\clearpage
\onecolumngrid

\renewcommand{\thesection}{\Roman{section}}

%Use the format (S#) for numbering equations in the supplementary material
\renewcommand{\theequation}{S\arabic{equation}}
\renewcommand{\thefigure}{S\arabic{figure}}
\renewcommand\bibnumfmt[1]{[S#1]}
\renewcommand{\citenumfont}{S}

%Section heading formatting
\setcounter{secnumdepth}{3}
\setcounter{equation}{0}\setcounter{figure}{0}
\renewcommand{\thesection}{S\Roman{section}}

\begin{center}
{\large \textbf{Supplementary Material for `Homodyne detection is optimal for quantum interferometry with path-entangled coherent states'}}\\\smallskip
Z. M. McIntyre and W. A. Coish\\
Department of Physics, McGill University, 3600 rue University, Montreal, QC, H3A 2T8, Canada
\end{center}

In this supplement, we provide additional results for quantum metrology with qubit--which-path (QWP) states. In Sec.~\ref{section:quantum-fisher}, we give a derivation of the quantum Fisher information of the QWP state accounting for photon loss and qubit dephasing. Next, in Sec.~\ref{sec:homodyning-scheme}, we calculate the classical Fisher information of the homodyning scheme presented in the main text, for a QWP state, and show that it is optimal in the ideal case of no photon loss or qubit dephasing. Finally, in Sec.~\ref{sec:photon-counting}, we show that in the absence of photon loss and qubit dephasing, photon counting followed by an $X$-basis qubit readout also constitutes an optimal strategy for a QWP state. We then derive the classical Fisher information for this counting-based strategy and show that, unlike the classical Fisher information of the homodyning scheme, it depends on the true value of the phase being estimated, similar to the case discussed in the main text for an ECS.

\section{Quantum Fisher information for qubit--which-path states}\label{section:quantum-fisher}

Here, we provide a detailed derivation of the quantum Fisher information $I_{\mathrm{Q}}(\rho_\phi,J_3)$ of QWP states with respect to $J_3=(a_1^\dagger a_1-a_2^\dagger a_2)/2$, accounting for photon loss and qubit dephasing.

The qubit and modes 1,2 of the interferometer are initially prepared in the state
\begin{equation}
    \ket{\mathrm{QWP}}=\frac{1}{\sqrt{2}}\left(\ket{\uparrow}\ket{\alpha,0}+\ket{\downarrow}\ket{0,\alpha}\right).
\end{equation}
The Hamiltonian describing a qubit undergoing dephasing due to a combination of classical noise and a quantum environment is given by
\begin{equation}
    H_\mathrm{QE}(t)=\frac{1}{2}[\omega_{\mathrm{q}}(t)+h]Z+H_{\mathrm{E}},
\end{equation}
where $Z=\ketbra{\uparrow}-\ketbra{\downarrow}$ is a Pauli-Z operator, $\omega_{\mathrm{q}}(t)$ is the randomly fluctuating qubit splitting, and $h$ is an operator acting on the qubit's environment (having decoupled Hamiltonian $H_{\mathrm{E}}$). As explained in the main text, we model photon loss in the interferometer via a beamsplitter-type unitary~\cite{zhang2013quantumS}
\begin{equation}\label{S:beamsplitter}
    R_{c,c_\ell}(p)=e^{\mathrm{arcsin}\sqrt{p}(c_\ell^\dagger c-\mathrm{h.c.})},
\end{equation}
describing scattering of photons from mode $c$ to loss mode $c_\ell$ with a per-photon probability $p$. Under the action of the lossy interferometer, the initial state $\rho(0)=\ketbra{\mathrm{QWP}}\otimes\ketbra{0}_\ell\otimes \rho_{\mathrm{E}}$ (where $\ket{0}_\ell$ denotes the common vacuum state of the loss modes and $\rho_{\mathrm
E}$ is the initial state of the environment) evolves into the mixed state
\begin{equation}\label{S:rho-phi}
    \rho_\phi(t)=\llangle\langle\mathrm{Tr}_\ell\{RU_\phi(t) \rho(0) U_\phi^\dagger(t) R^\dagger\}\rangle_{\mathrm{E}}\rrangle, \quad R=\prod_{i=1,2}R_{a_i,a_{\ell_i}}(p),
\end{equation}
where here, $\mathrm{Tr}_\ell$ denotes a trace over the state of the loss modes, $\langle\rangle_{\mathrm{E}}$ denotes an average over the initial state of the environment, $\llangle\rrangle$ is an average over realizations of $\omega_{\mathrm{q}}(t)$, and where evolution of the qubit and modes 1,2 is described by
\begin{equation}
    U_\phi(t)=\mathcal{T}e^{-i\int_0^t d\tau\:H_\mathrm{QE}(\tau)}\prod_{i=1,2}e^{-i\phi_i \hat{n}_i}.
\end{equation}
Here, $\mathcal{T}$ is the time-ordering operator and (as in the main text) $\phi_i$ is the phase acquired by photons passing through interferometer arm $i=1,2$, with number operator $n_i$. Note that the same state $\rho_\phi(t)$ is obtained regardless of the order in which $U_{\phi}(t)$ and $R$ are applied.

In terms of the eigenstates $\ket{\lambda_k}$ and eigenvalues $\lambda_k$ of $\rho_\phi$ [Eq.~\eqref{S:rho-phi}], the quantum Fisher information of $\rho_\phi$ with respect to $J_3$ is given by~\cite{toth2014quantumS} 
\begin{equation}\label{S:qfi-eigenstates}
    I_{\mathrm{Q}}(\rho_\phi,J_3)=2\sum_{k,j}\frac{(\lambda_k-\lambda_j)^2}{\lambda_k+\lambda_j}\lvert \langle \lambda_k\lvert J_3\rvert \lambda_j\rangle\rvert^2.
\end{equation}

To find the spectral decomposition of $\rho_\phi$, note that the action of the beamsplitter $R$ [cf.~Eq.~\eqref{S:rho-phi}] on $\ket{\alpha_1,\alpha_2}\ket{0}_{\ell}$ ($\ell=\ell_1,\ell_2$) is given by
\begin{align}\label{S:beamsplitter-action}
    R\ket{\alpha_1,\alpha_2}\ket{0}_{\ell}=\ket*{\sqrt{1-p}\alpha_1,\sqrt{1-p}\alpha_2}\ket{\sqrt{p}\alpha_1,\sqrt{p}\alpha_2}_{\ell_1,\ell_2}.
\end{align}
Using Eq.~\eqref{S:beamsplitter-action} and taking the trace over the loss modes, we then find that
\begin{equation}\label{S:diagonalize}
    \rho_\phi(t)=\frac{1}{2}\sum_{\sigma=\uparrow,\downarrow}\ketbra*{\Psi_\sigma}+\frac{1}{2}e^{-p\lvert\alpha\rvert^2-\chi(t)}\left[e^{-i\vartheta(t)}\ketbra*{\Psi_\uparrow}{\Psi_\downarrow}+\mathrm{h.c.}\right],
\end{equation}
where $\ket{\Psi_\uparrow}=\ket{\uparrow}\ket{e^{i\phi_1}\sqrt{1-p}\alpha,0}$ and $\ket{\Psi_\downarrow}=\ket{\downarrow}\ket{0,e^{i\phi_2}\sqrt{1-p}\alpha}$, and where $\chi(t)$ and $\vartheta(t)$ are defined via
\begin{align}\label{S:coherence-factor}
    &\llangle\langle U_\uparrow(t)U_\downarrow^\dagger(t)\rangle_{\mathrm{E}}\rrangle\equiv e^{-\chi(t)-i\vartheta(t)},\quad U_\sigma(t)=\mathcal{T}e^{-i\int_0^td\tau \langle\sigma\lvert H_\mathrm{QE}(\tau)\rvert \sigma\rangle}.
\end{align}
Closed-form expressions for $\chi(t)$ and $\vartheta(t)$ in terms of the spectral density of the environment can be found in a weak-coupling approximation and often (but not always) assuming approximate Gaussian fluctuations. Various forms can be found depending on details of the environmental initial conditions and the specific form of the qubit-environment coupling term $h$. Notably, $\vartheta(t)=\vartheta_\mathrm{dyn}(t)+\vartheta_\mathrm{q}(t)$ generally includes a contribution $\vartheta_\mathrm{q}(t)$ that is unique to a quantum environment~\cite{norris2016qubitS,
paz2017multiqubitS,
kwiatkowski2020influenceS,
mcintyre2022nonS}, in addition to the usual (classical) dynamical phase, $\vartheta_\mathrm{dyn}(t)=\int_0^t d\tau\left( \llangle\omega_q(\tau)\rrangle+\left<h(\tau)\right>_\mathrm{E}\right)$ (with $h(t)=e^{iH_\mathrm{E} t}he^{-iH_\mathrm{E} t}$). Expressions for $\vartheta_\mathrm{q}(t)$ given in, e.g., Refs.~\cite{norris2016qubitS,
paz2017multiqubitS,
kwiatkowski2020influenceS,
mcintyre2022nonS} in the context of dynamical decoupling can used in Eq.~\eqref{S:coherence-factor} by restricting to the case of free-induction decay (where no decoupling pulses are applied). 

Equation \eqref{S:diagonalize} is easy to diagonalize, giving eigenvalues and eigenvectors 
\begin{align}
    &\lambda_\pm(t)=\frac{1}{2}(1\pm e^{-p\lvert\alpha\rvert^2-\chi(t)}),\label{eigenval}\\
    &\ket{\lambda_\pm}_t=\frac{1}{\sqrt{2}}(\ket{\Psi_\downarrow}\pm e^{i\vartheta(t)}\ket{\Psi_\uparrow}).\label{eigenstate}
\end{align} 
Calculating the quantum Fisher information using Eq.~\eqref{S:qfi-eigenstates} then becomes a question of keeping track of indices, which run over $k=\pm$ as well as $k\neq \pm$ (for which $\lambda_k=0$). We break up the sum as follows:
\begin{equation}\label{S:qfi}
    I_{\mathrm{Q}}(\rho_\phi,J_3)=4\sum_{k=\pm}\sum_{j\neq \pm}\lambda_k\lvert\langle \lambda_j\lvert J_3\rvert\lambda_k\rangle\rvert^2+4(\lambda_+-\lambda_-)^2\lvert\langle \lambda_+\lvert J_3\rvert\lambda_-\rangle\rvert^2.
\end{equation}
The double sum in Eq.~\eqref{S:qfi} can be rewritten as
\begin{equation}
    \sum_{\substack{k=\pm\\j\neq \pm}}\lambda_k\lvert\langle \lambda_j\lvert J_3\rvert\lambda_k\rangle\rvert^2=\sum_{k=\pm}\lambda_k\langle\lambda_k\rvert J_3\sum_{j\neq \pm}\ketbra{\lambda_j}J_3\rvert\lambda_k\rangle,
\end{equation}
where $\sum_{j\neq\pm}\ketbra{\lambda_j}=\mathbbm{1}-\sum_{k=\pm}\ketbra{\lambda_k}$. From here, Eq.~\eqref{S:qfi} can be evaluated using $\lvert \langle\lambda_\pm\lvert J_3\rvert\lambda_\mp\rangle\rvert^2=\frac{1}{4}(1-p)^2\bar{n}^2$ and $\langle\lambda_\pm\lvert J_3^2\rvert\lambda_\pm\rangle=\frac{1}{4}[(1-p)^2\bar{n}^2+(1-p)\bar{n}]$, giving
\begin{equation}\label{S:quantum-fisher-QWP}
    I_{\mathrm{Q}}(\rho_\phi,J_3)=e^{-2p\bar{n}-2\chi(t)}(1-p)^2\bar{n}^2+(1-p)\bar{n},
\end{equation}
where $\bar{n}=\lvert\alpha\rvert^2$ is the average number of photons in the initial QWP state.

\section{Fisher information for homodyne detection followed by qubit readout}\label{sec:homodyning-scheme}

Next, we consider the measurement scheme proposed in the main text, consisting of (I) homodyne detection of the output modes $a_\pm$ of the interferometer with local-oscillator phases $\varphi_\pm$, followed by (II) measurement of the qubit in the $X$ basis $\ket{\pm}$, where $\ket{\pm}=(\ket{\uparrow}+\ket{\downarrow})/\sqrt{2}$. The local-oscillator phases are given in Eq.~(10) of the main text,
\begin{align}
    \begin{aligned}
        &\varphi_+=\frac{\pi}{2}+\bar{\phi},\\
        &\varphi_-=\bar{\phi},\\
        &\bar{\phi}=\frac{1}{2}(\phi_1+\phi_2).
    \end{aligned}
\end{align}

In order to calculate the classical Fisher information $I_{\mathrm{C}}(\phi)$ associated with Steps (I)-(II), we first need to evaluate the conditional probability distribution
\begin{align}\label{S:conditional-distribution}
    p(x\vert\phi)=\mathrm{Tr}\{U_{\mathrm{BS}}\rho_\phi U_{\mathrm{BS}}^\dagger \hat{\Pi}_x\},
\end{align}
where $\rho_\phi$ is given by Eq.~\eqref{S:diagonalize} and $U_{\mathrm{BS}}$ is a beamsplitter unitary that maps the annihilation operators $a_{i}$ ($i=1,2$) to new modes $a_\pm=(a_1\pm a_2)/\sqrt{2}$. In Eq.~\eqref{S:conditional-distribution}, we have also introduced the POVM element $\hat{\Pi}_x$ describing homodyne detection of modes $a_\pm$ with outcomes $x_\pm$ [Step (I)], followed by qubit readout with outcome $X=\pm$ [Step (II)]:
\begin{equation}
    \hat{\Pi}_{x}=\ketbra{X}\bigotimes_{\sigma=\pm} \ketbra{x_{\varphi_\sigma}},\quad \ket{x_{\varphi_\sigma}}=e^{-i\varphi_\sigma a_\sigma^\dagger a_\sigma}\ket{x_\sigma},
\end{equation}
where here, $\ket{x_\sigma}$ is an eigenstate of $\hat{x}_\sigma=(a_\sigma^\dagger+a_\sigma)/\sqrt{2}$ having eigenvalue $x_\sigma$.

Using Bayes' Rule, we can re-write the conditional distribution $p(x\vert\phi)$ as
\begin{equation}\label{S:bayes-rule}
p(x \lvert \phi)=p(X \lvert x_+,x_-, \phi)p(x_+,x_-\lvert \phi),    
\end{equation}
where, under the assumption that $\alpha\in\mathbb{R}$, the conditional probability distribution $p(x_+,x_-\vert\phi)$ governing the distribution of measurement outcomes for Step (I) is given by
\begin{equation}\label{S:probability-homodyne-QWP}
    p(x_+,x_-\vert \phi)=\prod_{s=\pm}g_s(x_s,\phi),
\end{equation}
where $g_s(x_s,\phi)=\pi^{-1/2}\exp{{-}[x_s-\mu_s(\phi)]^2}$, $\mu_+(\phi)=\sqrt{1-p}\alpha\sin{\frac{\phi}{2}}$, and $\mu_-(\phi)=\sqrt{1-p}\alpha\cos{\frac{\phi}{2}}$. In contrast to the case of the ECS [cf.~Eq.~(14) of the main text], the probability distribution governing the homodyne outcomes does not exhibit interference fringes. The relative phase giving rise to such interference is instead kicked back onto the state of the qubit, as we now show.

Using Eq.~\eqref{S:probability-homodyne-QWP}, we can evaluate the post-measurement state $\rho_{\textsc{q}}(x_+,x_-,\phi)$ of the qubit as
\begin{align}
    \rho_{\textsc{q}}(x_+,x_-,\phi)&=\frac{\mathrm{Tr}_{\mathrm{photons}}\{U_{\mathrm{BS}}\rho_\phi U_{\mathrm{BS}}^\dagger\bigotimes_{\sigma=\pm} \ketbra{x_{\varphi_\sigma}}\}}{p(x_+,x_-\vert\phi)}\\
    &=\frac{1}{2}\left[\ketbra{\uparrow}+\ketbra{\downarrow}+e^{-p\alpha^2-\chi(t)}\left(e^{-2i[x_+\mu_-(\phi)-x_-\mu_+(\phi)]-i\vartheta(t)}\ketbra{\uparrow}{\downarrow}+\mathrm{h.c.}\right)\right],
\end{align}
where, in the first equality, $\mathrm{Tr}_{\mathrm{photons}}$ denotes a partial trace over the state of the $a_+,a_-$ modes. The probability of obtaining a measurement outcome $\pm$ for the $X$-basis qubit measurement [Step (II)], conditioned on outcomes $x_+,x_-$ for the homodyne measurement, can then be evaluated as
\begin{align}
    p(\pm\vert x_+,x_-,\phi)&=\langle\pm \vert\rho_{\textsc{q}}(x_+,x_-,\phi)\vert\pm \rangle\\
    &=\frac{1}{2}\pm \frac{1}{2}e^{-p \alpha^2-\chi(t)}\mathrm{cos}\left[2x_+\mu_-(\phi)-2x_-\mu_+(\phi)+\vartheta(t)\right].
\end{align}

In the absence of photon loss ($p=0$) and qubit dephasing [$\chi(t)=\vartheta(t)=0$], the classical Fisher information for this measurement sequence is given by
\begin{equation}
    I_{\mathrm{C}}[p(x\vert\phi)]=\bar{n}^2+\bar{n},\quad p=0,\quad\chi=\vartheta=0,
\end{equation}
which is equal to the quantum Fisher information of the QWP state given in Eq.~(5) of the main text. When any of $p,\chi,\vartheta\neq 0$, the Fisher information can be evaluated numerically (Fig.~S1).

\begin{figure}
    \centering
    \includegraphics[width=0.5\textwidth]{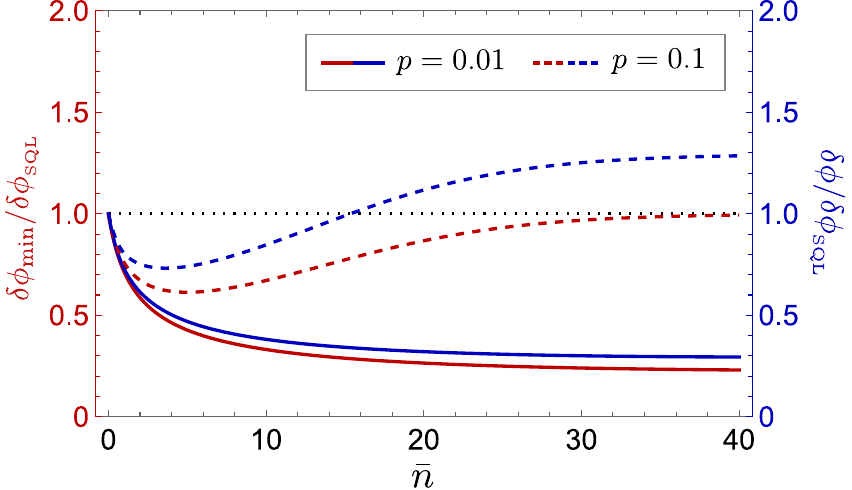}
    \caption{Left axis: Precision given by the quantum CRB $\delta\phi_{\mathrm{min}}$, relative to the standard quantum limit $\delta\phi_{\textsc{sql}}\equiv[(1-p)M\bar{n}]^{-1/2}$, for QWP states with  $p=0.01$ (solid red line) and $p=0.1$ (dashed red line). Right axis: Precision $\delta\phi$ that can be attained with homodyning and $X$-basis qubit readout for $p=0.01$ (solid blue line) and $p=0.1$ (dashed blue line). Here, we neglect qubit dephasing, $\chi=\vartheta=0$.}
    \label{fig:qwp-photon-loss}
\end{figure}

\section{Photon counting with QWP states}\label{sec:photon-counting}

\subsection{Photon counting saturates the quantum Cram\'er-Rao bound in the absence of photon loss}

In this section, we show that photon counting, supplemented by a final $X$-basis measurement of the qubit, can be used to saturate the quantum CRB in the absence of photon loss and qubit dephasing.  

For $p=0$, the state of the qubit and modes 1,2 just prior to the final beamsplitter $U_{\mathrm{BS}}$ is given by
\begin{equation}
    \ket{\mathrm{QWP}_\phi}=\frac{1}{\sqrt{2}}\left(\ket{\uparrow}\ket{e^{i\phi_1}\alpha,0}+\ket{\downarrow}\ket{0,e^{i\phi_2}\alpha}\right).
\end{equation}
Under the action of $U_{\mathrm{BS}}$, the modes $a_{1,2}$ transform according to $a_{1/2}=(a_+\pm a_-)/\sqrt{2}$, giving Eq.~(9) of the main text,
\begin{align}
\begin{aligned}
    \ket*{\widetilde{\mathrm{QWP}}_\phi}&=U_{\mathrm{BS}}\ket{\mathrm{QWP}_\phi}\\
    &=\frac{1}{\sqrt{2}}\left(\ket{\uparrow}\ket*{e^{i\phi_1}\tfrac{\alpha}{\sqrt{2}},e^{i\phi_1}\tfrac{\alpha}{\sqrt{2}}}+\ket{\downarrow}\ket*{{-}e^{i\phi_2}\tfrac{\alpha}{\sqrt{2}},e^{i\phi_2}\tfrac{\alpha}{\sqrt{2}}}\right).
\end{aligned}
\end{align}
Conditioned on detecting $m$ photons in mode $a_+$ and $n$ photons in mode $a_-$, the post-measurement state of the qubit is then given by
\begin{align}\label{S:post-meas}
\begin{aligned}
    \ket*{\mathrm{Q}_\phi^{m,n}}&=\frac{\langle m,n\vert \widetilde{\mathrm{QWP}}_\phi\rangle}{\sqrt{p(m,n)}}\\
    &=\frac{1}{\sqrt{2}}\left(e^{i\phi(m+n)}\ket{\uparrow}+e^{i\pi m}\ket{\downarrow}\right),
\end{aligned}
\end{align}
where here,
\begin{equation}\label{S:poisson}
    p(m,n)=\frac{e^{-\lvert \alpha\rvert^2}}{m!n!}\left(\frac{\lvert\alpha\rvert^{2}}{2}\right)^{m+n}
\end{equation}
is the probability of detecting $m$ and $n$ photons in modes $a_+$ and $a_-$, respectively. Due to the entanglement of the qubit with the electromagnetic field, the number-resolving measurements have the effect of kicking back a relative phase $\phi(m+n)$ onto the qubit superposition state [Eq.~\eqref{S:post-meas}]. The relative phase can then be estimated by measuring the qubit.

As in the homodyning scheme, we take the final measurement of the qubit to be a measurement in the Pauli-$X$ eigenbasis. Accounting also for the number-resolving measurements of modes $a_\pm$, the classical Fisher information $I_{\mathrm{C}}(\phi)$ for this strategy is given by 
\begin{equation}
    I_{\mathrm{C}}(\phi)=\sum_{X=\pm}\sum_{m,n=0}^\infty p(X,m,n\vert \phi)\left(\partial_\phi \mathrm{log}\:p(X,m,n\vert\phi)\right)^2.
\end{equation}
From Bayes' Rule, 
\begin{equation}
    p(X,m,n\vert \phi)=p_{m,n}(\pm\vert \phi)p(m,n),\label{S:bayes-rule-2}
\end{equation}
where $p(m,n)$ is the Poisson-distributed probability of detecting $m$ photons at one output and $n$ photons at the other [Eq.~\eqref{S:poisson}], and where 
\begin{align}
    p_{m,n}(\pm\vert \phi)&=\lvert\langle\pm\vert \mathrm{Q}_{\phi}^{m,n}\rangle\rvert^2\\&=\frac{1}{2}\left(1\pm \mathrm{cos}[\phi(m+n)-\pi m]\right)
\end{align} 
is the probability of measuring an eigenvalue $X=\pm 1$, given fixed values of $m$ and $n$. Since $p(m,n)$ is independent of $\phi$, it therefore follows that 
\begin{equation}\label{S:classical-fish}
    I_{\mathrm{C}}(\phi)=\sum_{m,n=0}^\infty p(m,n)I_{m,n}(\phi),
\end{equation}
where the $(m,n)$-conditioned classical Fisher information of the $X$-basis qubit measurement is given by
\begin{equation}\label{imn}
    I_{m,n}(\phi)=\sum_{X=\pm} \frac{[\partial_\phi p_{m,n}(X\vert \phi)]^2}{p_{m,n}(X\lvert \phi)}=(m+n)^2.
\end{equation}
Evaluating the average in Eq.~\eqref{S:classical-fish} then gives 
\begin{equation}
    I_{\mathrm{C}}(\phi)=\bar{n}^2+\bar{n},
\end{equation}
which is equal to the quantum Fisher information $I_{\mathrm{Q}}(\ket{\mathrm{QWP}_\phi})$ given in Eq.~(5) of the main text. Photon counting is therefore an optimal strategy in the absence of photon loss and qubit dephasing. 

\subsection{Classical Fisher information in the presence of photon loss}
\begin{figure}
    \centering
    \includegraphics[width=0.5\textwidth]{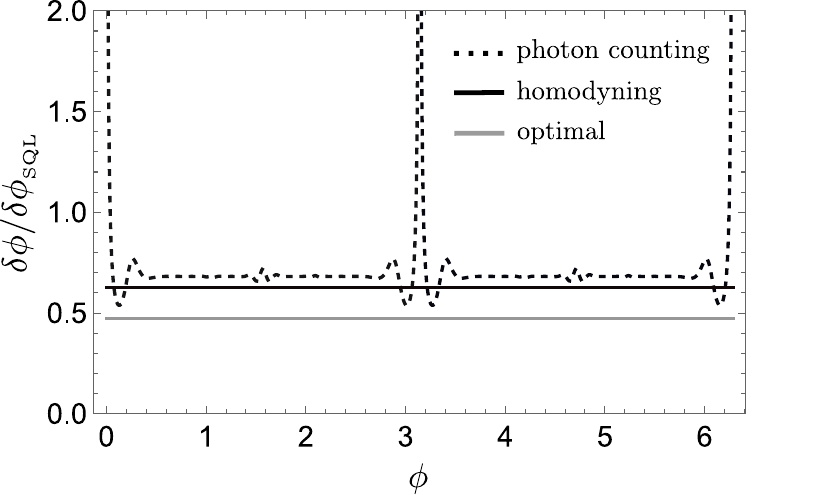}
    \caption{Precision $\delta\phi$ as a function of $\phi$, relative to the standard quantum limit $\delta\phi_{\textsc{sql}}\equiv [(1-p)M\bar{n}]^{-1/2}$, for a QWP state with $\bar{n}=10$ and $p=0.05$ ($5\%$ photon loss). The gray line indicating the performance of an optimal scheme is given by $\delta\phi_{\mathrm{min}}=1/\sqrt{M I_{\mathrm{Q}}}$, where $I_{\mathrm{Q}}$ is the quantum Fisher information given in Eq.~\eqref{S:quantum-fisher-QWP}. }
    \label{fig:phase-dep-QWP}
\end{figure}

Here, we neglect the effects of qubit dephasing as they do not affect the central message of this subsection concerning the $\phi$-dependence of the classical Fisher information associated with photon counting in the presence of photon loss. Dephasing can be incorporated with the replacements $e^{-p\lvert\alpha\rvert^2}\rightarrow e^{-p\lvert\alpha\rvert^2-\chi(t)}$ and $(m+n)\phi\rightarrow (m+n)\phi-\vartheta(t)$. 

In the presence of photon loss ($p\neq 0$), the post-measurement state of the qubit can be written as
\begin{align}
    \rho_{\textsc{q}}(m,n,\phi)&=\frac{\langle m,n\vert U_{\mathrm{BS}}\rho_\phi U_{\mathrm{BS}}^\dagger \vert m,n\rangle}{p(m,n)}\\
    &=\frac{1}{2}\left[\ketbra{\uparrow}+\ketbra{\downarrow}+e^{-p\lvert\alpha\rvert^2}\left(e^{i\phi(m+n)-im\pi}\ketbra{\uparrow}{\downarrow}+\mathrm{h.c.}\right)\right],
\end{align}
where $\rho_\phi$ is given by Eq.~\eqref{S:diagonalize} [with $\chi=\vartheta=0$], and where, in the first equality, the probability of detecting $m$ and $n$ photons in modes $a_+$ and $a_-$, respectively, is now given by
\begin{equation}\label{S:pmn-photon-loss}
    p(m,n)=\frac{e^{-(1-p)\lvert\alpha\rvert^2}}{m!n!}\left(\frac{(1-p)\lvert\alpha\rvert^2}{2}\right)^{m+n}.
\end{equation}
The probability $p_{m,n}(\pm\vert\phi)$ of measuring the qubit in state $\ket{\pm}$, conditioned on photon-counting outcomes $m$ and $n$, is then given by
\begin{equation}
    p_{m,n}(\pm\vert\phi)=\frac{1}{2}\left(1\pm e^{-p\lvert\alpha\rvert^2}\mathrm{cos}[\phi(m+n)-\pi m]\right).
\end{equation}

As in the $p=0$ case, Bayes' Rule [Eq.~\eqref{S:bayes-rule-2}] can be used to write the classical Fisher information in the form of Eq.~\eqref{S:classical-fish}, where for $p\neq 0$, the Fisher information of the qubit measurement is now given by [cf.~Eq.~\eqref{imn}] 
\begin{equation}\label{S:Imn-photon-loss}
    I_{m,n}(\phi)=\frac{(m+n)^2\mathrm{sin}^2(m+n)\phi}{e^{2p\lvert\alpha\rvert^2}-\mathrm{cos}^2(m+n)\phi}.
\end{equation}
Averaging Eq.~\eqref{S:Imn-photon-loss} over the distribution $p(m,n)$ given in Eq.~\eqref{S:pmn-photon-loss} gives the classical Fisher information $I(\phi)$. As was the case with the ECS, the Fisher information associated with the homodyning scheme is independent of the true value of the phase, while the Fisher information associated with photon counting vanishes for $\phi=0,\pi$, leading to divergences in $\delta\phi$ (Fig.~S2).

\end{document}